\documentclass[traditabstract]{aa}
\usepackage{graphicx}
\usepackage{amsmath,amssymb}
\usepackage{natbib}
\bibpunct{(}{)}{;}{a}{}{,}

\newcommand{\Teff}{T_\mathrm{eff}}
\newcommand{\nablaad}{\nabla_\mathrm{ad}}
\newcommand{\Jmb}{\dot J_\mathrm{mb}}

\def\msun{M_\odot}
\def\msol{M_\odot}
\def\mdot{\dot M}
\def\mjup{M_{\rm J}}
\def\msolyr{M_\odot \rm {yr}^{-1}}
\def\msun{M_\odot}

\def\nburst{N_{\rm burst}}
\def\tburst{\Delta t_{\rm burst}}

\begin{document}
   \title{Scenarios to explain extreme Be depletion in solar-like stars: accretion or rotation effects?}
  \author{M. Viallet, I. Baraffe}

   \institute{Physics and Astronomy, University of Exeter, Stocker Road, Exeter, UK EX4 4QL\\
   \email{mviallet@astro.ex.ac.uk, i.baraffe@exeter.ac.uk} }

   \date{Received; accepted}

  \abstract{ 
Studies of beryllium abundance in large samples of solar-type stars show a small fraction of extremely beryllium-deficient stars, which challenges our current understanding of light element depletion in these stars. We suggest two possible scenarios that may explain this high level of Be depletion: early accretion and rotational mixing. We show that in both cases, the conditions required to reach the observed level of Be depletion are quite extreme, which explains the very small fraction of detected Be outliers. We suggest that substantial Be depletion can be obtained in stars if they were fast rotators in the past, with high initial rotational velocities and short disc lifetimes. Our analysis suggests that rotational mixing may not be efficient enough to deplete Be in less than 10 Myr. Consequently, the detection of strongly Be-deficient stars in clusters younger than $\sim$ 10 Myr may provide a genuine signature of accretion process and the proof that some protostars may undergo many extreme bursts of accretion during their embedded phases of evolution.}

     
  \titlerunning{Extreme Be depletion in solar-type stars}
  \authorrunning{M. Viallet \& I. Baraffe}
   
 \maketitle

\section{Introduction}
\label{intro}

The light chemical elements lithium, beryllium and boron provide good tracers of  internal mixing processes in stellar interiors. A huge amount of observational and theoretical works has been devoted to the analysis of their abundances at the surface of various types of stars. Interest in these elements has recently increased within the context of exoplanet discoveries because of possible links between Li depletion in a star and the presence of a planet orbiting this star (see e.g Sousa et al. 2010 and references therein). This suggestion motivated extensive studies of solar-type stars with the main goal of analysing Li and Be in planet hosting stars and of confirming or refuting the above-mentioned link \citep{takeda_beryllium_2011, galvez-ortiz_beryllium_2011, 2012ApJ...746...47D}. This has yielded large samples of homogeneous data, which provide excellent  opportunities to study properties and trends of light-element depletion. One striking property resulting from these studies is a small fraction of highly Be-deficient stars in sufficiently large samples, which questions our current understanding of Be destruction in solar-type and low-mass stars. 

In a sample of 82 stars, \cite{2004A&A...425.1013S} found three abnormally Be-depleted objects, namely \object{HD 4391}, \object{HD 20766} and \object{HD 280807}. These Be-deficient stars are dwarfs with $\Teff$ in the range 5700 - 5900 K. Lithium depletion is poorly constrained but is also significant, with Li abundances  A(Li)\footnote{The notation A(E) defines the abundance of element E and is equal to  $\log[N(E)/N(H)]$ + 12, where N(E) and N(H) are  the mole fraction of element E and of hydrogen, respectively.} $<$ 1 (upper limit). According to the measured $v \sin i$ of these stars, their rotation rate is similar to that of the Sun, and their estimated ages are between 1 and 2 Gyr. \cite{takeda_beryllium_2011} found four strongly Be-depleted objects within their sample of 118 stars: \object{HIP 17336}, \object{HIP 32673}, \object{HIP 64150}, and \object{HIP 75676}. These objects have similar properties as those of the \cite{2004A&A...425.1013S} sample: strong Li depletion, an age of $\sim 3-5$ Gyr, and low $v \sin i \sim $ 2 km/s. The properties of these seven objects are summarised in Table \ref{tab:stars}. Existing calculations for Be depletion in this mass range, based on rotational mixing scenarios \citep{pinsonneault_evolutionary_1989} or mixing by internal waves \citep{montalban_mixing_2000}, cannot explain these low Be abundances (see e.g Fig. 4 of  \citealt{2012ApJ...746...47D}). 

In this paper, we explore two scenarios that may explain these Be-deficient outliers, which are observed in an effective temperature range $5000-6000$ K and correspond to a mass range $0.8-1.1\ \msol$. 
The first scenario  is based on episodic accretion at early stages of evolution  (\S \ref{sect:accretion}), which can strongly affect the internal structure of young accreting objects and enhance the depletion of light elements, following the ideas of \cite{2009ApJ...702L..27B} and \cite{baraffe_effect_2010}. The second scenario is based on rotational mixing. We adopt a simplified treatment of rotational mixing in stellar interiors, which we describe and validate in \S \ref{sect:rotation}. We quantitatively show (\S \ref{rot:results}) that substantial Be depletion can be obtained in stars if they were fast rotators in the past. This idea was mentioned in \cite{takeda_beryllium_2011}, but no quantitative studies exist to confirm it. Discussion and conclusion follow in Sect. \ref{sect:conclusion}.
 
\begin{table}[t]
   \caption{Parameters of strongly Be-depleted objects.}
   \label{tab:stars}
   \centering
   \begin{tabular}{l c c c c c} % Column formatting, @{} suppresses leading/trailing space
     \hline \hline
      Star & $ \Teff$ &  A(Be) & A(Li) & Age & $v \sin i$  \\
             &      (K)    &           &          &  (Gyr) &  (km/s)   \\
      \hline
      HD 4391$^{(a)}$    & 5878  &     0.64    &  $<1.09$ & 1.2 & 2.72 \\
      HD 20766$^{(a)}$  & 5733  & $<-0.09$ &  $<0.97$ & 1.6 & 1.98 \\
      HD 20807$^{(a)}$  & 5843  &     0.36    &  $<1.07$ & 2.1 & 1.74 \\
      HIP 17336$^{(b)}$ & 5671  & $<-0.85$ &  $<0.8$  & 3.89 & 2.07 \\
      HIP 32673$^{(b)}$ & 5724  & $<-0.78$ &  $<1.0$  & 3.16 & 2.83 \\
      HIP 64150$^{(b)}$ & 5800  & $<-0.88$ &  $<1.0$  & 4.2 & 2.21 \\
      HIP 75676 $^{(b)}$ & 5772  & $<-0.96$ & $<1.1$  & 4.16 & 2.18 \\
      \hline
   \end{tabular}
  \tablefoot{
      \tablefoottext{a}{From \cite{2004A&A...425.1013S}}       
      \tablefoottext{b}{From  \cite{takeda_behavior_2010} and \cite{takeda_beryllium_2011}.}      
      }
\end{table}

\section{Episodic accretion scenario}
\label{sect:accretion}

\begin{figure}[t] 
   \centering
%   \includegraphics[height=7cm,width=7cm]{be_accretion} 
%    \vspace{-1cm}
   \includegraphics[width=0.9\linewidth,trim=0 150 0 80,clip=true]{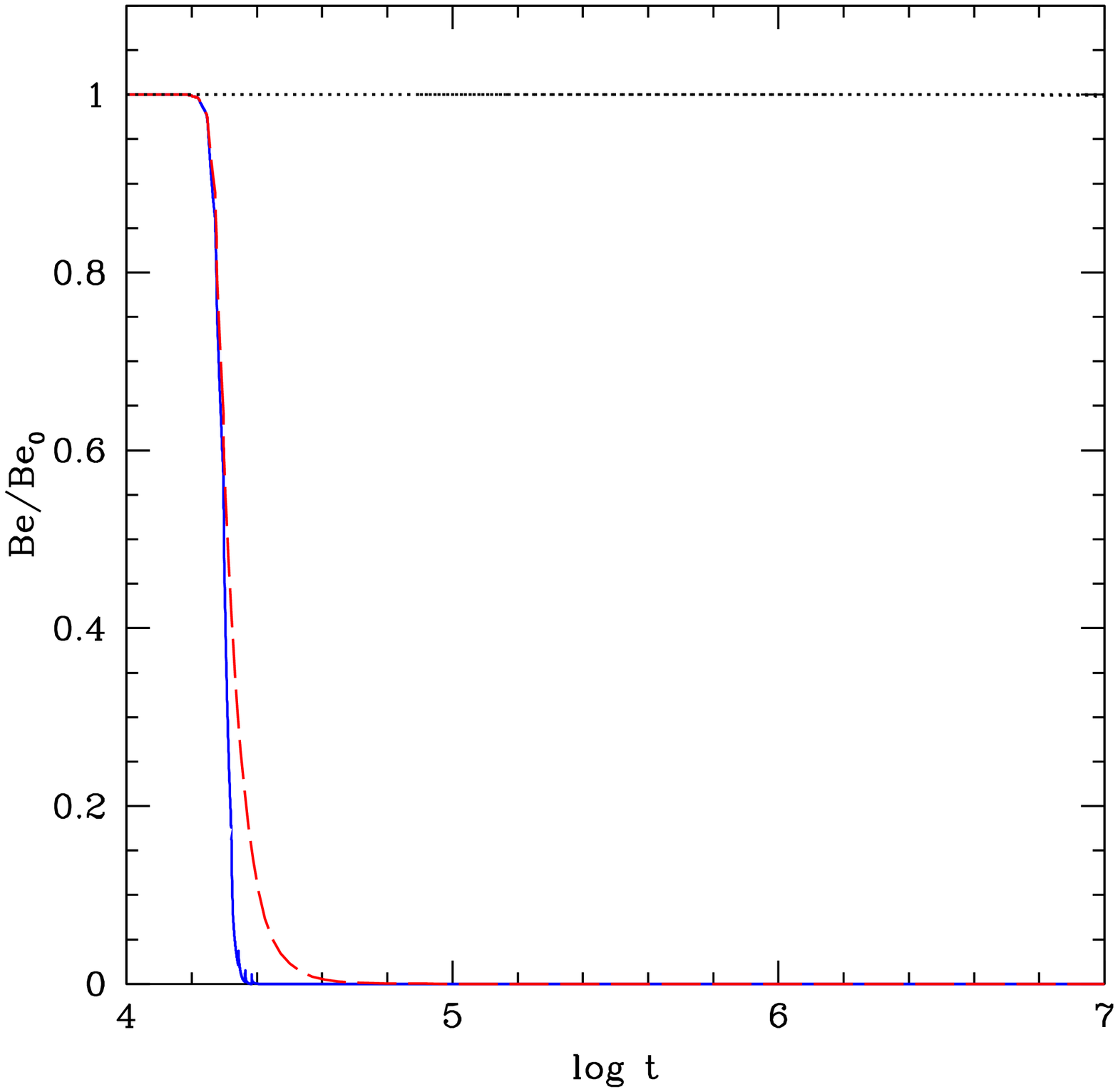} 
   \caption{Abundance of Be (divided by the initial abundance Be$_0$) versus time (in yr) in stars that have undergone a given number $\nburst$ of accretion bursts with $\mdot$=5 10$^{-4}\ \msolyr$ and duration $\tburst$ = 100 yr. The blue solid line corresponds to an accreting sequence with final mass 1 $\msol$ with $\nburst$=20 bursts. The red dashed line corresponds to a final mass 0.8 $\msun$ with $\nburst$=16. The accretion process starts from an initial seed mass of 10 $\mjup$ (see Baraffe \& Chabrier 2010 for details). The black dotted line indicates the Be abundance for non-accreting models of 0.8 $\msol$ or 1 $\msol$, which are the same for  both masses since Be is not depleted in those models. 
   }
   \label{fig:accretion}
\end{figure}

\cite{2009ApJ...702L..27B} and \cite{baraffe_effect_2010} recently investigated the effect of episodic accretion at very early stages of evolution on the structure and Li depletion of low-mass objects. 
\cite{2009ApJ...702L..27B} showed that bursts of accretion can produce objects significantly more compact than their non-accreting counterpart of same mass and age. These authors suggested that non-steady accretion during the embedded phase of protostar evolution produces the observed spread in luminosity in the Herzsprung-Russel diagram (HRD) of young clusters.  As a consequence of the more compact structure, \cite{baraffe_effect_2010} showed that objects undergoing very strong accretion bursts have significantly higher central temperatures and can be severely Li-depleted. This scenario provides an explanation for the unexpected level of Li depletion observed in a few low-mass members of young clusters ($\sim$ a few Myr). 
Based on the same calculations as in \cite{baraffe_effect_2010}, we find that significant Be depletion can take place in stars that have undergone short and very intense bursts of accretion with  $\mdot \sim$  5 10$^{-4}\ \msolyr$. Those results are illustrated in Fig. \ref{fig:accretion}, which displays the abundance of Be as a function of time at early stages of evolution of low-/solar-mass stars. 
These calculations show that the early history of accretion can significantly deplete Be in stars in the mass range of interest. Note that  Li is totally destroyed in the models displayed in Fig. \ref{fig:accretion}. As explained in \cite{baraffe_effect_2010}, the more compact and hotter structure of accreting models increases the maximum temperature reached at the bottom of the convective envelope, which increases the level of Li and Be destruction compared to the non-accreting counterparts of same mass. We find that substantial depletion of Be only happens if the protostar undergoes several bursts (more than ten) of extreme intensities with rates $\gg 10^{-4}\ \msolyr$. According to  numerical simulations of collapsing cloud cores, typical mass accretion rates onto the protostar vary in the range 10$^{-6}\ \msolyr - 10^{-4}\ \msolyr$ \citep{2010ApJ...719.1896V, 2010ApJ...724.1006M}. Several burst events with rates exceeding a few times 10$^{-4}\ \msolyr$ up to $\sim$ 10$^{-3}\ \msolyr$ are predicted, but only for high cloud masses with high initial angular momentum values (see \citealt{2010ApJ...719.1896V}).
This suggests that high levels of Be depletion by accretion process should only be produced under quite extreme conditions and should therefore be rare events. 

\section{Rotational mixing scenario: the model}
\label{sect:rotation}

In the following, we explore quantitatively whether rotation  can explain extreme Be depletion, based on the implementation of rotation effects in the same stellar evolution code as in the previous section to study accretion effects (see Baraffe et al. 2009 and references therein).  Different parametrisations for the treatment of rotational effects in 1D stellar evolution codes exists in the literature, see e.g. \cite{pinsonneault_evolutionary_1989}; \cite{heger_presupernova_2000}; \cite{denissenkov_physical_2000}; \cite{maeder_evolution_2010} and references therein. These formalisms are usually based on the treatment of \cite{endal_evolution_1978} \citep[e.g.][]{pinsonneault_evolutionary_1989} or on the model of \cite{zahn_circulation_1992} and its subsequent developments \citep[see e.g.][]{maeder_stellar_1998}. 
 All these formalisms contain free parameters that are calibrated to reproduce observations (e.g solar surface rotational velocity, surface abundances of light elements) and  yield very different results when applied to the same case study, as recently illustrated by \cite{2012MNRAS.419..748P}.
Strong uncertainties still remain regarding the treatment of rotation and its effect on stellar structure and evolution, despite very sophisticated and complicated formalisms.
As our aim here is to explore a possible connection between rotation and extreme depletion of light elements, we chose to follow a simple approach to explore the effects of rotation on Li/Be depletion.

\subsection{Rotation law}

Our approach adopts the hypothesis of \cite{1997A&A...326.1023B}, namely:

\begin{enumerate}
\item \emph{Solid body rotation} - Instead of solving an angular momentum equation, we assume solid-body rotation. The underlying hypothesis is that angular momentum is efficiently transported within the interior. The physical processes transporting angular momentum can be of magnetic origin (e.g. magnetic torque) and/or hydrodynamic origin (shear turbulence, large-scale currents, internal waves). The efficiencies and relative contributions of these processes to the total angular momentum transport are still poorly understood. Note that helioseismology shows that the bulk of the solar radiative zone is in nearly solid-body rotation, see e.g. \cite{2007Sci...316.1591G}.
\item \emph{Disc locking} - In the early stages of evolution, the star is efficiently coupled to its accretion disc and the rotation rate $\Omega$ remains constant during the disc lifetime: $\Omega(t \le \tau_d) = \Omega_0$.
\item \emph{Magnetic breaking} - Angular momentum is lost through magnetic breaking by the stellar wind. The rate of loss is computed using  Kawaler's law \citep{kawaler_angular_1988}, including saturation of the losses when $\Omega > \Omega_\mathrm{sat}$. We use the parametrisation

\begin{equation}
\Jmb = - K \Omega^3 \Big( \frac{R}{R_\odot} \Big)^{1/2}  \Big( \frac{M}{M_\odot} \Big)^{-1/2} \mathrm{\ \ \ if\ } \Omega < \Omega_\mathrm{sat}
\end{equation}

\noindent and

\begin{equation}
\Jmb = - K \Omega_\mathrm{sat}^2 \Omega \Big( \frac{R}{R_\odot} \Big)^{1/2}  \Big( \frac{M}{M_\odot} \Big)^{-1/2} \mathrm{otherwise.}%\mathrm{\ \ \ if\ } \Omega \ge \Omega_\mathrm{sat}
\end{equation}

We use \noindent $\Omega_\mathrm{sat}$=14 $\Omega_\odot$ \citep{1997A&A...326.1023B}. $K$ is a constant that is calibrated to yield the correct surface velocity $v \sim 2$ km/s of a solar model at $t=4.6$ Gyr. This yields a value $K = 2.7\times 10^{47}$, in agreement with previous works \citep[see e.g.][]{1997A&A...326.1023B}. Magnetic breaking is applied from the end of the disc locking phase.

\end{enumerate}

Within this approach, the rotational history of a star is entirely determined by two free parameters: $\Omega_0$ and $\tau_d$. Values adopted for these parameters are discussed in Sect. \ref{rot:params}.

\subsection{Rotational mixing of chemical species}

Assuming that transport of angular momentum is efficient enough to maintain solid-body rotation, we only need to  treat rotational mixing of chemical species. Rotational mixing in stably stratified regions driven by dynamical flows (e.g. meridional circulation, shear instabilities) is implemented in stellar evolution models to explain various observational features (see e.g Maeder \& Meynet 2010 and references therein). 
It is taken into account by introducing an effective ``turbulent'' diffusivity $D_t$ in the evolution equation of chemical species,

\begin{equation}
\rho \frac{\partial X_l}{\partial t} = \frac{1}{r^2} \frac{\partial }{\partial r} \big( r^2 D_\mathrm{t} \rho  \frac{\partial X_l}{\partial r} \big).
\end{equation}

\begin{figure}[t] 
%\vspace{-4cm}
   \centering
   \includegraphics[width=\linewidth]{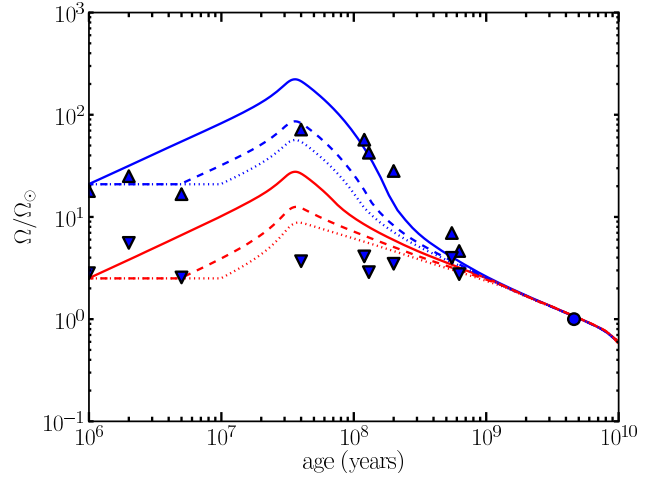}    
   \caption{Evolution of the rotation rate for an initially slow (lower red curves) and  fast (upper blue curves) rotating solar-mass star, for  different disc lifetimes: $\tau_d = 1$ Myr (solid curves);  $\tau_d = 5$ Myr (dashed curves); $\tau_d = 10$ Myr (dotted curves). The solar value is indicated by a solid circle. Direct and inverted triangles are observational data compiled in Table 1 of \cite{bouvier_lithium_2008} and represent respectively the 10th and 75th percentiles of observed rotational period distribution.}
   \label{fig:rotation_rate}
\end{figure}

\begin{figure*}[t]
   \centering
%   \vspace{-3cm}
   \includegraphics[width=\linewidth]{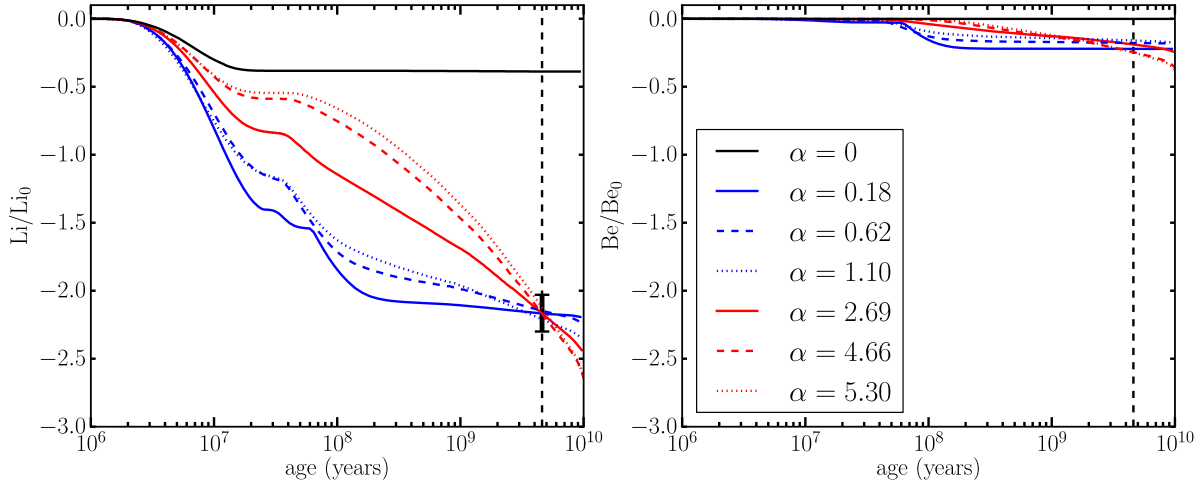} 
   \caption{Calibration of $\alpha$ for different rotational histories (represented with the same colour-coding as in Fig. \ref{fig:rotation_rate}) of a 1 M$_\odot$ star using solar values for the ratio of present to initial abundances of Li (Li/Li$_0$=1/140, see left panel). The black error bar shows the range $100 \lesssim$ d(Li) $\lesssim 200$. The right panel shows the corresponding depletion of beryllium and the calibrated values of $\alpha$. In both panels, the case without rotational mixing is shown for reference (black continuous line).}
   \label{fig:calibration}
\end{figure*}

The expression for $D_t$ significantly varies from one author to the other (see references above and also Potter et al. 2012).
We chose to adopt a simple prescription  \citep[see e.g.][]{2004SoPh..220..243R} that will permit us to capture the essential effects of rotational mixing:

\begin{equation}
D_t = \alpha r | U(r) |,
\label{dt}
\end{equation}

\noindent where $\alpha$ is a free parameter and $U(r)$ is related to the radial velocity component of the meridional circulation:

\begin{equation}
u_r (r, \theta) = U(r) P_2(\cos \theta),
\end{equation}

\noindent where $P_2$ is the second Legendre polynomial. In the solid-body rotation hypothesis, $U(r)$ is given by
(Richard et al. 2004)

\begin{equation}
U(r) = \frac{P}{\rho g T c_p} \frac{L}{M_\star} \frac{1}{\nablaad - \nabla} E_\Omega,
\end{equation}

\noindent where $E_\Omega$ is defined by

\begin{equation}
E_\Omega = 2 \big( 1 - \frac{\Omega^2}{2\pi G \rho} \big ) \frac{\tilde{g}}{g} = \frac{8}{3}\frac{\Omega^2 r^3}{G M}\big( 1 - \frac{\Omega^2}{2\pi G \rho} \big ),
\end{equation}

\noindent and $M_\star$ is given by

\begin{equation}
M_\star = M(r) \big ( 1 - \frac{\Omega^2}{2\pi G \rho_\mathrm{m}} \big )
\end{equation}

\noindent with $\rho_\mathrm{m}$  the mean density within the sphere of radius $r$.
 
We neglect the effect of molecular weight gradients, which have stabilising effects and consequently decrease the efficiency of mixing. Molecular weight gradients appear in stellar interiors when a significant amount of central hydrogen has been burned during the main sequence evolution. Since most of the rotationally induced depletion occurs earlier, i.e. for $t < 0.1 $ Gyr, before any significant $\mu$-gradient appears in the interior (see \S \ref{rot:params}), neglecting $\mu$-gradient effects in our analysis will not change the conclusions of this work. 

This model is simple to implement in a stellar evolution code and has the advantage of involving only one free parameter $\alpha$ that needs to be calibrated (see \S \ref{rot:params}). We stress that in this model, the rotation history is totally decoupled from the mixing process, i.e. $\alpha$ has no feedback on $\Omega(t)$.

\subsection{Rotational histories and calibration of $\alpha$}
\label{rot:params}

We adopted different rotational histories  based on the work of \cite{bouvier_lithium_2008}. We used three values for the disc lifetime $\tau_d$: 1 Myr, 5 Myr and 10 Myr. These values are  representative of  disc fraction analysis in young clusters of different ages, with a disc fraction exponentially decaying between $\sim$ 1 and $\sim$ 10 Myr \citep{2009AIPC.1158....3M}.  As in \cite{bouvier_lithium_2008}, we constructed rotational models for slow and fast rotators, with initial rotation periods $P_0$ equal to 10 days (2.5 $\Omega_\odot$) and 1.2 days ($\sim 21\ \Omega_\odot$), respectively. 

The different values of $\tau_d$ and $P_0$ mentioned above define six different rotational histories. Figure \ref{fig:rotation_rate} shows the evolution of the rotation rate for each case. Observational data compiled in Table 1 of \cite{bouvier_lithium_2008} are shown for comparison. The models cannot reproduce the lower envelope of observed rotational periods, failing at predicting sufficiently low rotation periods  at $t\sim 40$ Myr (see Fig. \ref{fig:rotation_rate}). This may stem from the solid-body rotation assumption. \cite{bouvier_lithium_2008} was able to reproduce the lower envelope of the distribution by assuming strong decoupling between the core and the envelope. An envelope rotating ten times slower than the core yields the observed low surface velocities. On the other hand, \cite{bouvier_lithium_2008} showed that strong coupling between core and envelope, resulting in nearly solid-body rotation, reproduces the upper envelope of the velocity distribution. Therefore, solid-body rotation assumption is a good proxy for fast rotating models during the main-sequence (it is not excluded, however, that significant differential rotation can develop during the PMS, see \citealt{2012A&A...539A..70E}).

The calibration of the free parameter $\alpha$ defining the diffusion coefficient $D_t$ (see Eq. \ref{dt}) is based on the observed solar depletion of lithium, taken to be Li/Li$_0$ = 1/140 \citep[see][]{1996A&A...312.1000R}. A 1 M$_\odot$ star model is evolved with the six different rotational histories defined above and for each cases, $\alpha$ is calibrated to obtain the observed Li depletion at the solar age ($t_\odot = 4.6$ Gyr). The result of this procedure is shown in Fig. \ref{fig:calibration}. It shows that beryllium is also depleted, with d(Be) $\sim -0.2 $ dex at the solar age. Note that recent measurements of abundances in the solar photosphere do not suggest a significant depletion in beryllium \citep[see][]{1998SSRv...85..161G, 2009ARA&A..47..481A}.
The calibration process yields values of $\alpha$ in a range between 0.18 and 5.30, as summarised in Fig. \ref{fig:calibration}. Higher values of $\alpha$ are needed for lower rotation rates to reproduce the expected lithium depletion. The left panel of Fig. \ref{fig:calibration} shows that the depletion history of lithium is quite different for fast rotators and slow rotators: fast rotators have essentially depleted their lithium after $t \sim 100$ Myr and the depletion on the main-sequence is modest due to the low values of $\alpha$, whereas slow rotators undergo a more steady depletion due to higher values of $\alpha$. As expected, in our approach depletion increases with increasing $\alpha$ and increasing rotation rates.

\begin{figure*}[t] 
   \centering
%    \vspace{-4cm}
%   \includegraphics[angle=90, height=15cm,width=15cm]{A_vs_Teff} 
   \includegraphics[width=\linewidth]{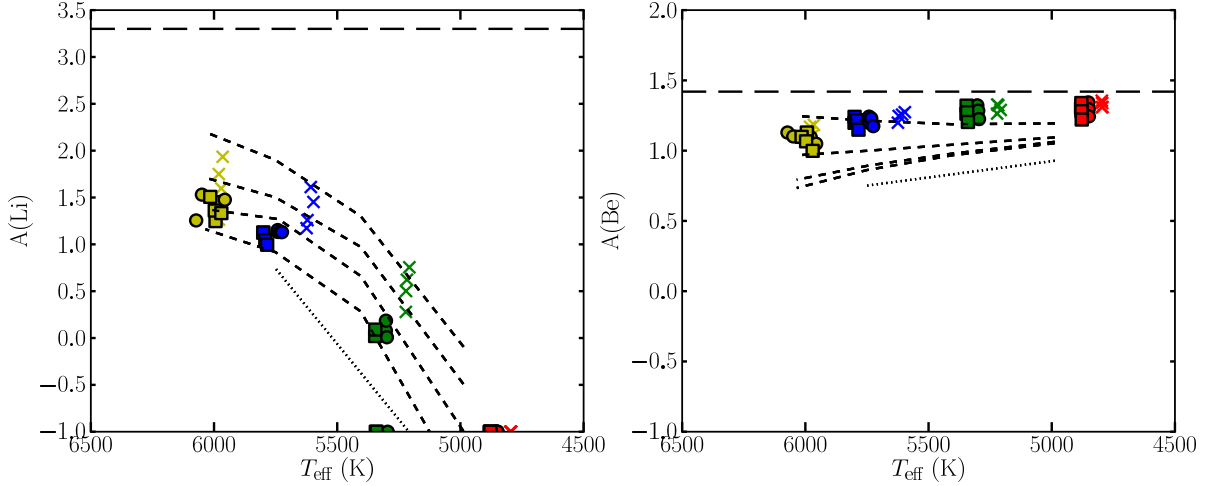} 
%   \vspace{-3cm}
   \caption{Lithium (left) and beryllium (right) depletion versus effective temperature for four different masses $M=$ 0.8 $M_\odot$ (red symbols), 0.9 $M_\odot$ (green symbols), 1 $M_\odot$ (blue symbols), 1.1 $M_\odot$ (yellow symbols) at three different ages t=1.7 Gyr (crosses), 4.6 Gyr (dots), 6 Gyr (squares). For each mass, at a given age, the results of the six models corresponding to different rotational histories are displayed (see text). Initial abundances are shown as horizontal long-dashed lines, lithium abundances below -1 are plotted at -1. The four dashed lines correspond to models A of \cite{pinsonneault_rotation_1990}. The dotted curve corresponds to model C3 of \cite{pinsonneault_rotation_1990}, which predicts the highest Be depletion.}
   \label{fig:AvsTeff}
\end{figure*}

\subsection{Validation of the model for different stellar masses/ages}

To additionally test the validity of our model assumptions, we explored the behaviour of the calibration obtained in \S \ref{rot:params} for different masses in the range $0.8 - 1.1 M_\odot$. Each stellar mass was evolved with the six adopted rotational histories (see Fig. \ref{fig:rotation_rate}) and their corresponding value of $\alpha$ defined in \S \ref{rot:params} (see right panel of Fig. \ref{fig:calibration}). 
Results are displayed in Fig. \ref{fig:AvsTeff} \footnote{Initial abundances of Li and Be were set to the meteoritic values A(Li)=3.3 and A(Be)=1.42 \citep[see][]{1998SSRv...85..161G}} at three different ages: t=1.7 Gyr, 4.6 Gyr and 6 Gyr. The models predict a decreasing abundance of Li with $\Teff$; below 5500 K, all  models have A(Li) $<$ 1, which is roughly the detection limit. This is in qualitative agreement with observations, see e.g. Fig. 5 of \cite{galvez-ortiz_beryllium_2011} and Fig. 5 of \cite{takeda_behavior_2010}. For a given mass, the spread of symbols at $t=1.7$ Gyr over $\sim 2$ dex reflects the spread seen in the left panel of Fig. \ref{fig:calibration}.

Beryllium abundances show a much lower dispersion, because Be depletion is much less sensitive to rotational mixing and age. Our models show a slight increase of A(Be) with decreasing $\Teff$ for the mass range considered here. The reason is that for decreasing stellar mass, the convective envelope reaches deeper layers and rotational mixing becomes less and less efficient as a result of the decrease in $r|U(r)|$ and hence in the mixing diffusivity $D_t$ (see Eq. \ref{dt}).

Different authors disagree on the observed trend of A(Be) with $\Teff$. Figure 9 of \cite{2009ApJ...691.1412B} suggests a significant \emph{increase} of beryllium abundance with decreasing $\Teff$ in the range $\Teff \sim 5500 - 6500$ K; Fig. 6 of \cite{2012ApJ...746...47D} (see also Fig. 4 of \citealt{galvez-ortiz_beryllium_2011}) exhibits a clear trend of \emph{decreasing} beryllium abundance with decreasing $\Teff$ in the range $\Teff =  5500 - 6200$ K; whereas Fig. 9 of \cite{takeda_beryllium_2011} exhibits a \emph{constant} beryllium abundance in the range $\Teff \sim 5500 - 6000$~K. Given these opposite results, we cannot conclude on the validity of our approach to predict the observed trend of A(Be) with $\Teff$.

We show in Fig. \ref{fig:AvsTeff} models of \cite{pinsonneault_rotation_1990} for the depletion of Li and Be at an age of $1.7$ Gyr, used by \cite{galvez-ortiz_beryllium_2011} and \cite{2012ApJ...746...47D} for comparison with their observations. These models agree very well with our calibration of the depletion of lithium, whereas the right panel of Fig. \ref{fig:AvsTeff} shows that our parametrisation yields a lower Be depletion than that of \cite{pinsonneault_rotation_1990}.

\section{Rotational mixing scenario: results}
\label{rot:results}

\begin{figure}[t] 
   \centering
   \includegraphics[width=\linewidth]{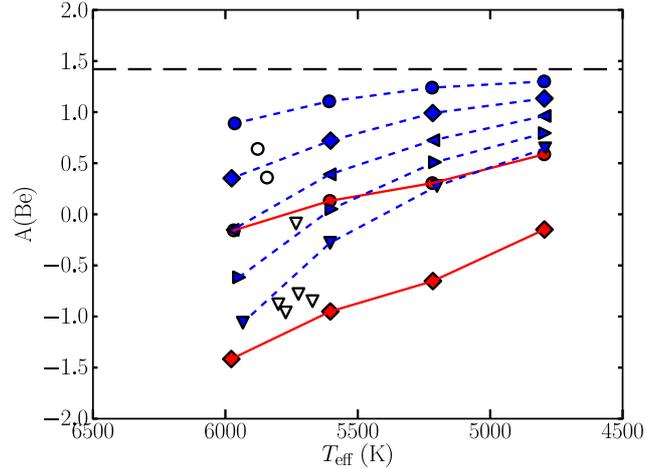} 
   \caption{Beryllium depletion at $t=1.7$ Gyr for our fast-rotating model ($P_0 = 1.2$ d) with $\tau_d = 1$ Myr (continuous red lines) and $\tau_d = 5$ Myr (dashed blue lines), for all four masses $M=0.8, 0.9, 1, 1.1\ M_\odot$. The symbols correspond to different values of $\alpha $: 1 (dot), 2 (diamond), 3 (left triangle), 4 (right triangle), 5 (down triangle). Open symbols are the objects of Table \ref{tab:stars}, triangles stand for upper limits. The initial Be abundance is shown as a horizontal dashed line. }
   \label{fig:depletion}
\end{figure}

\begin{figure*}[t] 
   \centering
   \includegraphics[width=\linewidth]{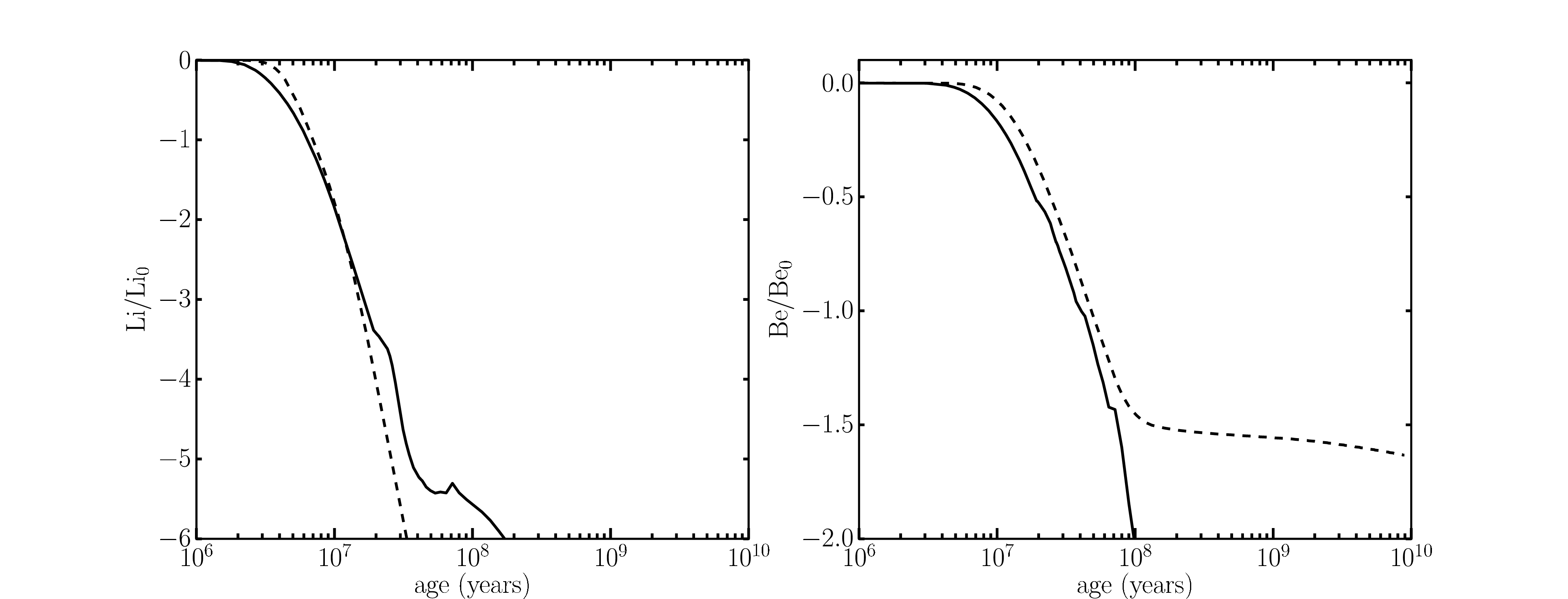} 
   \caption{Abundances of Li (left panel) and Be (right panel), divided by initial abundance, versus time for the fast rotator model with  $\tau_d = 1$ Myr and $P_0$ = 1.2 d, with $\alpha$=2. Results are displayed for $M=0.8\ \msol$ (dashed line) and $M=1.1\ \msol$ (solid line).}
   \label{fig:tlibe}
\end{figure*}

The simplified treatment of rotational mixing adopted in this work has been validated in the previous section. Calibrating the free parameter $\alpha$ based on the Sun's properties and on a given rotational history yields values for $\alpha$ in the range $\sim 0.2 - 5.3$. This calibration provides realistic predictions for Li and, to a lesser extent, Be depletion as a function of mass (or $\Teff$) in the mass range of interest $0.8-1.1\ \msol$. The model can therefore be used to explore conditions required for extreme Be depletion, adopting the range of calibrated $\alpha$ values but for other rotational histories, to explain the objects listed in Table \ref{tab:stars}. 
These have common properties: low values of $v \sin i$, roughly at the present day solar rotation rate (2 km/s), and an estimated age older than 1 Gyr. As shown in Fig. \ref{fig:rotation_rate}, different rotational histories yield models that converge towards the same track after $\sim$ 1 Gyr. Consequently, observed Be-deficient stars could have been faster rotators than the Sun in the past, even though they now have low $v \sin i$. Note that we ignore the rotational history of the Sun and whether it was a slow or fast rotator in the past. 

Given this unknown, we assumed two rotational histories that produce the fastest rotators at early ages, ($\tau_d = 1$ Myr, $P_0$ = 1.2 d) and ($\tau_d = 5$ Myr, $P_0$ = 1.2 d), and adopted higher values of $\alpha$, but within the calibration range. 
The resulting Be depletion at $t=1.7$ Gyr \footnote{Results for ages $\gtrsim 1$ Gyr are similar.} is shown in Fig. \ref{fig:depletion}. The objects of Table \ref{tab:stars} are also shown for comparison. The figure shows that for the fastest rotating model ($\tau_d = 1$ Myr, $P_0$ = 1.2 d), a value of $\alpha = 2$ yields strong depletion (A(Be)$< -1$) for 1 $M_\odot$ and 1.1 $M_\odot$ stars, i.e. in the $\Teff$ range of the observed outliers. For a slightly slower rotating model ($\tau_d = 5$ Myr, $P_0$ = 1.2 d), higher values of $\alpha$ are required to reach a similar level of Be depletion, but still within the calibration range. Slower rotational histories were not explored since they would require values of $\alpha$ outside the calibration range.
Interestingly enough, significant beryllium depletion (A(Be) $\sim -1$ or $\log$ Be/Be$_0 \, \sim -2$) is difficult to achieve for $M < 1\ \msol$ with our model, except for unreasonable values of $\alpha$ outside the calibration range. 

Variations of Li and Be abundance versus time in the fastest rotating model ($\tau_d = 1$ Myr, $P_0$ = 1.2 d) with $\alpha$=2 are
displayed in Fig. \ref{fig:tlibe} for the two extreme cases M=$0.8\ \msol$ and M=$1.1\ \msol$. 
For these fast rotators, Li is rapidly depleted, decreasing by a factor 5 after 5 Myr and a factor 100 after 10 Myr, whereas Be depletion starts later in those models, at ages $>$ 10 Myr. Even with the highest value of $\alpha$ within the calibration range ($\alpha = 5$), Be is depleted by only a factor two within the first 10 Myr. 

\section{Conclusion}
\label{sect:conclusion}

We have presented two possible scenarios that can explain extremely Be-deficient solar-type stars. The first scenario is based on the effect of accretion at very early stages of evolution of the protostar. For very intense bursts, we showed that Be can be very rapidly destroyed, producing strongly depleted objects at ages of only a few Myr, which could explain the Be-deficient solar-type stars listed in Table \ref{tab:stars}. High levels of Be depletion, as observed,  can only be reached if the protostar undergoes several bursts (more than ten) of extreme intensities ($\mdot \gg 10^{-4}\ \msolyr$). Those events are predicted by numerical simulations of cloud collapse (see e.g \citealt{2010ApJ...719.1896V}), but only for massive and initially rapidly rotating clouds. This scenario should therefore produce only very few Be-deficient stars, which agrees with the few observed numbers (typically 3-4 objects in a sample of $\sim$ 80-100 stars, see \S \ref{intro}).  

The other possible scenario is based on rotational mixing. We have explored those effects based on a simplified treatment of rotation in a 1D stellar evolution code. Rotational mixing was described in our approach by a single free parameter. Our model assumed solid-body rotation throughout the stellar evolution of both slow and rapid rotators. Slow rotators are thought to develop a high degree of differential rotation \citep[see][]{bouvier_lithium_2008} and it is not excluded that rapid rotators can develop significant differential rotation during the pre-main sequence phase \citep[see][]{2012A&A...539A..70E}. Differential rotation induces shear mixing, which is by definition absent from our model. This is embedded in our single free parameter $\alpha$, calibrated to reproduce the solar lithium depletion. We showed quantitatively that a model assuming solid-body rotation that also has values of the free parameter within the calibration range can reach the observed level of Be-depletion assuming initially fast rotators with short disc lifetimes ($\tau_d \sim$ 1 Myr). 
The observed level of Be depletion in the above-mentioned outliers requires initially very fast rotators with very short coupling times between the star and its disc. From a statistical point of view, those objects should belong to the upper envelope of the observed rotational period distribution as a function of age and should thus only represent a small fraction of objects ($<$ 10\%, see e.g \citealt{bouvier_lithium_2008}). This could be consistent with the small fraction of observed Be-deficient solar-type stars. 
This exploratory study highlights the possible connection between rotation and extreme depletion of light elements. Given our assumptions, we cannot rule out the possibility for slow rotators, which are thought to develop a high degree of differential rotation, to reach extreme depletion through an efficient mixing by shear turbulence. Additionnal studies should be devoted to investigate this possibility and confirm our findings. 
Ideally, these should be based on improved modelling of rotational mixing, including a consistent picture of angular momentum transport and turbulent mixing in stellar interiors. Improvement is expected from the development of multi-D numerical simulations of stellar interiors, based either on existing numerical tools \citep{espinosa_lara_dynamics_2007,2007ApJ...669.1190B}, or on new tools (see e.g \citealt{2011A&A...531A..86V}). 
Multi-D numerical simulations will likely yield a better calibration of the free parameters inherent to current rotation formalisms used in 1D stellar evolution codes and provide better understanding of the different mechanisms leading to angular momentum transport and chemical mixing in stars. 

If accretion and rotation are two possible processes that can produce Be-deficient stars, it may be difficult to find different observational signatures that would allow distinguishing one process from the other. 
Regarding a possible signature of accretion process, our analysis suggests that rotation is hardly able to deplete Be within the first $\sim$ 10 Myr of evolution of solar-type stars. It may be argued that differential rotation could enhance the depletion in this early phase through additional mixing by shear instabilities \citep[see][]{2012A&A...539A..70E}. In our 1.1 $\msol$ model, the radiative core develops at $t \sim 1.4 $ Myr, and Be-burning starts at the centre of the star at $t \sim 3-4 $ Myr. Given the size of the radiative core, we estimate that the mixing diffusivity should be enhanced by more than one order of magnitude, reaching values of about $10^7$ cm$^2$/s, to alter the abundance in the convective envelope, and thus at the surface, in less than $t=10$ Myr. Although this depends on how strong differential rotation can develop during this early phase, it seems unlikely to reach such high diffusivities. Detection of strongly Be-deficient stars in clusters younger than $\sim$~10 Myr could thus provide the genuine signature of the accretion process and the proof that protostars may undergo many extreme bursts of accretion during their embedded phases. The search for Be-deficient outliers in young clusters, but also in older environments to improve their statistics, is therefore worth pursuing since their existence may improve our understanding of fundamental processes in stellar formation and evolution, such as accretion processes, rotational mixing, and transport of angular momentum.

\begin{acknowledgement}
MV acknowledges support from a Newton International Fellowship from the Royal Society. IB thanks the support of the European Research Council under the European Community's Seventh Framework Programme (FP7/2007-2013 Grant Agreement No. 247060) and the Royal Society award WM090065, which funded part of this work. The authors acknowledge useful discussions with Cyril Georgy, Brian Chaboyer, and Garik Israelian. The authors also thank the anonymous referee for his/her comments that helped to improve the quality of the paper.
\end{acknowledgement}

\bibliographystyle{aa}
\bibliography{biblio}

\begin{thebibliography}{33}
\expandafter\ifx\csname natexlab\endcsname\relax\def\natexlab#1{#1}\fi

\bibitem[{{Asplund} {et~al.}(2009){Asplund}, {Grevesse}, {Sauval}, \&
  {Scott}}]{2009ARA&A..47..481A}
{Asplund}, M., {Grevesse}, N., {Sauval}, A.~J., \& {Scott}, P. 2009, \araa, 47,
  481

\bibitem[{{Ballot} {et~al.}(2007){Ballot}, {Brun}, \&
  {Turck-Chi{\`e}ze}}]{2007ApJ...669.1190B}
{Ballot}, J., {Brun}, A.~S., \& {Turck-Chi{\`e}ze}, S. 2007, \apj, 669, 1190

\bibitem[{Baraffe \& Chabrier(2010)}]{baraffe_effect_2010}
Baraffe, I. \& Chabrier, G. 2010, Astronomy and Astrophysics, 521, 44

\bibitem[{{Baraffe} {et~al.}(2009){Baraffe}, {Chabrier}, \&
  {Gallardo}}]{2009ApJ...702L..27B}
{Baraffe}, I., {Chabrier}, G., \& {Gallardo}, J. 2009, \apjl, 702, L27

\bibitem[{{Boesgaard} \& {Krugler Hollek}(2009)}]{2009ApJ...691.1412B}
{Boesgaard}, A.~M. \& {Krugler Hollek}, J. 2009, \apj, 691, 1412

\bibitem[{Bouvier(2008)}]{bouvier_lithium_2008}
Bouvier, J. 2008, Astronomy and Astrophysics, 489, L53

\bibitem[{{Bouvier} {et~al.}(1997){Bouvier}, {Forestini}, \&
  {Allain}}]{1997A&A...326.1023B}
{Bouvier}, J., {Forestini}, M., \& {Allain}, S. 1997, \aap, 326, 1023

\bibitem[{{Delgado Mena} {et~al.}(2012){Delgado Mena}, {Israelian},
  {Gonz{\'a}lez Hern{\'a}ndez}, {Santos}, \& {Rebolo}}]{2012ApJ...746...47D}
{Delgado Mena}, E., {Israelian}, G., {Gonz{\'a}lez Hern{\'a}ndez}, J.~I.,
  {Santos}, N.~C., \& {Rebolo}, R. 2012, \apj, 746, 47

\bibitem[{Denissenkov \& Tout(2000)}]{denissenkov_physical_2000}
Denissenkov, P.~A. \& Tout, C.~A. 2000, Monthly Notices of the Royal
  Astronomical Society, 316, 395

\bibitem[{{Eggenberger} {et~al.}(2012){Eggenberger}, {Haemmerl{\'e}}, {Meynet},
  \& {Maeder}}]{2012A&A...539A..70E}
{Eggenberger}, P., {Haemmerl{\'e}}, L., {Meynet}, G., \& {Maeder}, A. 2012,
  \aap, 539, A70

\bibitem[{Endal \& Sofia(1978)}]{endal_evolution_1978}
Endal, A.~S. \& Sofia, S. 1978, The Astrophysical Journal, 220, 279

\bibitem[{Espinosa~Lara \& Rieutord(2007)}]{espinosa_lara_dynamics_2007}
Espinosa~Lara, F. \& Rieutord, M. 2007, Astronomy and Astrophysics, 470, 1013

\bibitem[{{G{\'a}lvez-Ortiz} {et~al.}(2011){G{\'a}lvez-Ortiz}, {Delgado-Mena},
  Gonz{\'a}lez~Hern{\'a}ndez, Israelian, Santos, Rebolo, \&
  Ecuvillon}]{galvez-ortiz_beryllium_2011}
{G{\'a}lvez-Ortiz}, M.~C., {Delgado-Mena}, E., Gonz{\'a}lez~Hern{\'a}ndez,
  J.~I., {et~al.} 2011, Astronomy and Astrophysics, 530, 66

\bibitem[{{Garc{\'{\i}}a} {et~al.}(2007){Garc{\'{\i}}a}, {Turck-Chi{\`e}ze},
  {Jim{\'e}nez-Reyes}, {Ballot}, {Pall{\'e}}, {Eff-Darwich}, {Mathur}, \&
  {Provost}}]{2007Sci...316.1591G}
{Garc{\'{\i}}a}, R.~A., {Turck-Chi{\`e}ze}, S., {Jim{\'e}nez-Reyes}, S.~J.,
  {et~al.} 2007, Science, 316, 1591

\bibitem[{{Grevesse} \& {Sauval}(1998)}]{1998SSRv...85..161G}
{Grevesse}, N. \& {Sauval}, A.~J. 1998, \ssr, 85, 161

\bibitem[{Heger {et~al.}(2000)Heger, Langer, \&
  Woosley}]{heger_presupernova_2000}
Heger, A., Langer, N., \& Woosley, S.~E. 2000, The Astrophysical Journal, 528,
  368

\bibitem[{Kawaler(1988)}]{kawaler_angular_1988}
Kawaler, S.~D. 1988, The Astrophysical Journal, 333, 236

\bibitem[{{Machida} {et~al.}(2010){Machida}, {Inutsuka}, \&
  {Matsumoto}}]{2010ApJ...724.1006M}
{Machida}, M.~N., {Inutsuka}, S.-i., \& {Matsumoto}, T. 2010, \apj, 724, 1006

\bibitem[{Maeder \& Meynet(2010)}]{maeder_evolution_2010}
Maeder, A. \& Meynet, G. 2010, New Astronomy Reviews, 54, 32

\bibitem[{Maeder \& Zahn(1998)}]{maeder_stellar_1998}
Maeder, A. \& Zahn, J. 1998, Astronomy and Astrophysics, 334, 1000

\bibitem[{{Mamajek}(2009)}]{2009AIPC.1158....3M}
{Mamajek}, E.~E. 2009, in American Institute of Physics Conference Series, Vol.
  1158, American Institute of Physics Conference Series, ed. {T.~Usuda,
  M.~Tamura, \& M.~Ishii}, 3--10

\bibitem[{Montalb{\'a}n \& Schatzman(2000)}]{montalban_mixing_2000}
Montalb{\'a}n, J. \& Schatzman, E. 2000, Astronomy and Astrophysics, 354, 943

\bibitem[{Pinsonneault {et~al.}(1990)Pinsonneault, Kawaler, \&
  Demarque}]{pinsonneault_rotation_1990}
Pinsonneault, M.~H., Kawaler, S.~D., \& Demarque, P. 1990, The Astrophysical
  Journal Supplement Series, 74, 501

\bibitem[{Pinsonneault {et~al.}(1989)Pinsonneault, Kawaler, Sofia, \&
  Demarque}]{pinsonneault_evolutionary_1989}
Pinsonneault, M.~H., Kawaler, S.~D., Sofia, S., \& Demarque, P. 1989, The
  Astrophysical Journal, 338, 424

\bibitem[{{Potter} {et~al.}(2012){Potter}, {Tout}, \&
  {Eldridge}}]{2012MNRAS.419..748P}
{Potter}, A.~T., {Tout}, C.~A., \& {Eldridge}, J.~J. 2012, \mnras, 419, 748

\bibitem[{{Richard} {et~al.}(2004){Richard}, {Th{\'e}ado}, \&
  {Vauclair}}]{2004SoPh..220..243R}
{Richard}, O., {Th{\'e}ado}, S., \& {Vauclair}, S. 2004, \solphys, 220, 243

\bibitem[{{Richard} {et~al.}(1996){Richard}, {Vauclair}, {Charbonnel}, \&
  {Dziembowski}}]{1996A&A...312.1000R}
{Richard}, O., {Vauclair}, S., {Charbonnel}, C., \& {Dziembowski}, W.~A. 1996,
  \aap, 312, 1000

\bibitem[{{Santos} {et~al.}(2004){Santos}, {Israelian}, {Randich},
  {Garc{\'{\i}}a L{\'o}pez}, \& {Rebolo}}]{2004A&A...425.1013S}
{Santos}, N.~C., {Israelian}, G., {Randich}, S., {Garc{\'{\i}}a L{\'o}pez},
  R.~J., \& {Rebolo}, R. 2004, \aap, 425, 1013

\bibitem[{Takeda {et~al.}(2010)Takeda, Honda, Kawanomoto, Ando, \&
  Sakurai}]{takeda_behavior_2010}
Takeda, Y., Honda, S., Kawanomoto, S., Ando, H., \& Sakurai, T. 2010, Astronomy
  and Astrophysics, 515, 93

\bibitem[{Takeda {et~al.}(2011)Takeda, Tajitsu, Honda, Kawanomoto, Ando, \&
  Sakurai}]{takeda_beryllium_2011}
Takeda, Y., Tajitsu, A., Honda, S., {et~al.} 2011, Publications of the
  Astronomical Society of Japan, 63, 697

\bibitem[{{Viallet} {et~al.}(2011){Viallet}, {Baraffe}, \&
  {Walder}}]{2011A&A...531A..86V}
{Viallet}, M., {Baraffe}, I., \& {Walder}, R. 2011, \aap, 531, A86

\bibitem[{{Vorobyov} \& {Basu}(2010)}]{2010ApJ...719.1896V}
{Vorobyov}, E.~I. \& {Basu}, S. 2010, \apj, 719, 1896

\bibitem[{Zahn(1992)}]{zahn_circulation_1992}
Zahn, J. 1992, Astronomy and Astrophysics, 265, 115

\end{thebibliography}

%\begin{thebibliography}{}
%\bibitem[]{} Sousa, S., Fernandes, J., Israelian, G., Santos, N. 2010, \aap, 512, L5
%\end{thebibliography}{}

\end{document}